# Neighbor Contrastive Learning on Learnable Graph Augmentation

Xiao Shen[1], Dewang Sun[1], Shirui Pan[2], Xi Zhou[1], Laurence T. Yang[1,3]

[1] Hainan University
[2] Griffith University
[3] St. Francis Xavier University

shenxiaocam@163.com, dwsun@hainanu.edu.cn, s.pan@griffith.edu.au, xzhou@hainanu.edu.cn, ltyang@hainanu.edu.cn

## Abstract

Recent years, graph contrastive learning (GCL), which aims to learn representations from unlabeled graphs, has made great progress. However, the existing GCL methods mostly adopt human-designed graph augmentations, which are sensitive to various graph datasets. In addition, the contrastive losses originally developed in computer vision have been directly applied to graph data, where the neighboring nodes are regarded as negatives and consequently pushed far apart from the anchor. However, this is contradictory with the homophily assumption of networks that connected nodes often belong to the same class and should be close to each other. In this work, we propose an end-to-end automatic GCL method, named NCLA to apply neighbor contrastive learning on learnable graph augmentation. Several graph augmented views with adaptive topology are automatically learned by the multi-head graph attention mechanism, which can be compatible with various graph datasets without prior domain knowledge. In addition, a neighbor contrastive loss is devised to allow multiple positives per anchor by taking network topology as the supervised signals. Both augmentations and embeddings are learned end-to-end in the proposed NCLA. Extensive experiments on the benchmark datasets demonstrate that NCLA yields the state-of-the-art node classification performance on self-supervised GCL and even exceeds the supervised ones, when the labels are extremely limited. Our code is released at https://github.com/shenxiao-cam/NCLA.

## Introduction

Over the past several years, graph neural networks (GNNs) have attracted great attention due to their outstanding performance in various graph mining tasks, such as node classification (Kipf and Welling 2017), link prediction (Shen and Chung 2020) and graph classification (Hamilton, Ying, and Leskovec 2017). Most existing GNNs are trained in a supervised manner which heavily relies on a large amount of well-annotated labels. However, in the real-world applications, it is often resource-expensive and time-consuming to collect abundant labeled graph structured data (Shen et al. 2020a; Shen et al. 2020b; Shen, Mao, and Chung 2020; Wu et al. 2020; Dai et al. 2022).

Contrastive learning (CL) is one of the most representative self-supervised learning techniques that can reduce the reliance on manual labels. CL has demonstrated unprecedented performance on unsupervised representation learning in computer vision (CV) (Zhu et al. 2020) and natural language processing (NLP) (Aberdam et al. 2021). Inspired by the development of CL, recently, tremendous endeavors have been devoted to graph contrastive learning (GCL) (Hassani and Khasahmadi 2020; Zhu et al. 2020; Xia et al. 2022a; Xia et al. 2022b; Zheng et al. 2022), which couple GNNs with CL to learn robust representations from unlabeled graphs.

Most existing GCL methods adopt a similar paradigm. Firstly, they employ various graph augmentation strategies, such as node dropping (You et al. 2020), edge perturbation (Zhu et al. 2020), attribute masking (Zhu et al. 2021), subgraph (Yang et al. 2022) and graph diffusion (Hassani and Khasahmadi 2020), to generate several graph augmented views with discrepancy. Secondly, they apply the contrastive losses widely utilized in CV, such as InfoNCE (Van den Oord, Li, and Vinyals 2018), normalized temperature-scaled cross-entropy (NT-Xent) (Zhu et al. 2020), Jensen-Shannon Divergence (JSD) (Nowozin, Cseke, and Tomioka 2016) and Triplet loss (Schroff, Kalenichenko, and Philbin 2015) to extract the common core information between different augmented views, according to the InfoMax (Linsker 1988) principle. Despite the prosperous development of GCL, there are some drawbacks in the standard paradigm, in terms of graph augmentations and contrastive objectives.

The theoretical and empirical analysis in CL showed that good augmented views should be diverse while keeping the task-relevant information intact (Tian et al. 2020). However, the existing handcraft graph augmentation strategies, which randomly perturb graph topology, would fail to keep the task-relevant information intact. For example, dropping an important edge can heavily damage the graph topology that are highly related to downstream tasks and consequently cause the low quality of graph embeddings (Zhu et al. 2021). In addition, owing to the diverse nature of graph data, there is no universal graph augmentation suitable for different datasets (You et al. 2020; You et al. 2021). Thus, the existing ad-hoc graph augmentations have to be manually chosen for each graph dataset based on prior domain knowledge or trial-and-errors (You et al. 2020), which significantly limit both efficiency and general applicability of existing GCL methods.

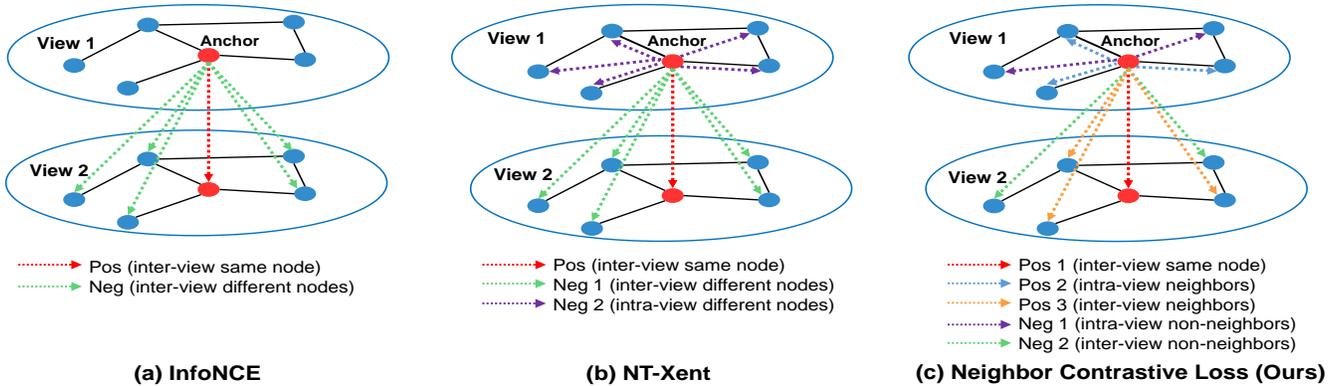

Figure 1. The comparisons of positive and negative pairs defined in the three node-node contrastive losses, i.e., InfoNCE, NT-Xent and our proposed neighbor contrastive loss. The red nodes denote the anchor in view 1 and the same node in view 2. The full lines in black denote the original edges in the network. The dotted lines with arrows in different colors denote the positive and negative pairs formed with the anchor.

On the other hand, existing GCL methods directly apply the contrastive losses originally proposed in CV to graph data (Qiu et al. 2020; You et al. 2020; Zhu et al. 2020; Wan et al. 2021a; Wan et al. 2021b; Zhu et al. 2021), while paying no attention to the inherent distinction between images and graphs. The contrastive losses are utilized to guide the representation learning to pull positive pairs together and push negative pairs far apart. As shown in Figure 1(a) and 1(b), in both InfoNCE and NT-Xent, a single positive pair is formed with each anchor, by creating different augmented views of the same node. Then, InfoNCE regards all the other different nodes from different view as negatives. While NT-Xent introduces more negatives, where all different nodes within a view and from different view are regarded as negatives. It is worth noting that in both InfoNCE and NT-Xent, the neighboring nodes are regarded as negatives and then pushed apart from the anchor. However, in GCL, the GNNs coupled with CL are generally based on the homophily[1] assumption that connected nodes often belong to the same class (McPherson, Smith-Lovin, and Cook 2001). In other words, connected nodes should be similar to each other rather than far apart. Due to the inherent distinction between images and graphs, directly applying the contrastive losses developed in CV to GCL would overlook the network topology and result in the embeddings contradict with the homophily assumption of GNNs.

To remedy the aforementioned limitations, in this work, we propose a new GCL method, named NCLA, which applies neighbor contrastive learning on learnable graph augmentation. On one hand, NCLA adopts the multi-head graph attention network (GAT) (Veličković et al. 2018) to generate $K$ learnable graph augmented views with adaptive topology. Such learnable augmentation can be automatically compatible with various graph datasets without prior domain knowledge. In addition, in contrast to inappropriate handcraft graph augmentations which might heavily damage the original topology, the attention-based learnable augmented views generated by NCLA would keep exactly the same nodes and edges as the original graph but with different adaptive edge weights. Moreover, unlike the existing GCL methods which utilize exactly the same GNN encoder with the tied learnable parameters for different augmented views (You et al. 2020; Zhu et al. 2020; Zhu et al. 2021), in NCLA, each augmented view has its own learnable parameters. As a result, NCLA can generate safer graph augmentation without improper modifications of original topology while still guaranteeing the diversity between different augmented views. On the other hand, unlike previous GCL methods which directly utilize the contrastive losses originally proposed in CV (e.g., InfoNCE or NT-Xent), we devise a new neighbor contrastive loss for node-node GCL. The proposed neighbor contrastive loss is a novel extension to the NT-Xent loss (Zhu et al. 2020), by taking network topology as the supervised signals to define positives and negatives in GCL. Specifically, instead of only forming a single positive pair per anchor as in NT-Xent, the proposed neighbor contrastive loss allows for multiple positives per anchor. Such multiple positives are drawn from the same node in different view and also the neighbors of the anchor within a view and from different view, as illustrated in Figure 1(c). Consequently, the non-neighbors of the anchor within a view and from different view would be regarded as the intra-view and inter-view negatives. The contributions of this work can be summarized as follows:

1) As opposed to most existing GCL methods which have to manually pick handcraft graph augmentations per dataset, the proposed NCLA is the first to adopt the multi-head graph attention mechanism as the learnable graph augmentation function, where each head corresponds to one augmented view. Such attention-based learnable graph augmentation avoids improper modification of the original topology and can be automatically compatible with various graph datasets.

2) Existing GCL methods directly apply the contrastive losses in CV to graph data which overlooks network topology. To the best of our knowledge, our work makes one of the pioneering attempts to study neighbor contrastive learning in node-node GCL, which allows for multiple positives per anchor by taking network topology as the supervised signals.

3) In the standard GCL paradigm, graph augmentation and embedding learning are conducted in two stages which might require bi-level optimization. In contrast, in NCLA, graph augmentation is learned along with embeddings end-to-end, which yields high flexibility and ease of use.

4) The extensive experiments on various graph datasets demonstrate that NCLA consistently outperforms the state-of-the-art GCL methods or even some supervised GNNs on semi-supervised node classification with scarce labels.

---
[1] Studying graphs going beyond homophily is out of scope of this work.

# Related Work

In line with the focus of our work, we briefly review existing GCL methods from two aspects, i.e., graph augmentation and contrastive objective.

## Graph Augmentation

In the typical GCL framework, the first step is to generate graph augmented views with discrepancy by various graph augmentations (Ding et al. 2022). For example, DGI (Velickovic et al. 2019) augments the original graph by row-wise shuffling node attributes. GRACE (Zhu et al. 2020) corrupts graph by removing edges and masking attributes uniformly. To improve GRACE, GCA (Zhu et al. 2021) removes edges and masks attributes adaptively, by assigning different probabilities based on the centrality heuristics. MVGRL (Hassani and Khasahmadi 2020) augments the input graph by graph diffusion and generates both local and global structural views. GraphCL (You et al. 2020) proposes four types of graph augmentations, including node dropping, edge perturbation, attribute masking and subgraph. However, the aforementioned handcraft augmentations have been shown to be sensitive to different graph datasets, owing to the diverse nature of graphs, thus limiting the efficiency and generalizability of the GCL methods (You et al. 2020; Lee, Lee, and Park 2022). To avoid manually tuning dataset-specific graph augmentations, the concurrent work (Lee, Lee, and Park 2022; Mo et al. 2022; Wang et al. 2022; Xia et al. 2022a) propose to remove augmentations in GCL.

The proposed NCLA generates thoroughly learnable graph augmentation by the multi-head graph attention mechanism. Such attention-based learnable augmentation can be automatically compatible with various graph datasets, and also avoid improper modification of the original graph topology.

## Contrastive Objective

Two common contrastive modes in GCL are node-graph and node-node (Liu et al. 2022). The node-graph GCL methods contrast node-level representations with graph-level representation. For example, DGI (Velickovic et al. 2019) contrasts the node representations of the original and corrupted graph with the original graph representation. MVGRL (Hassani and Khasahmadi 2020) contrasts node representations of one view with graph representation of the other view. On the other hand, the node-node GCL methods contrast node-level representations between positive and negative node pairs. For example, GMI (Peng et al. 2020) contrasts the input neighborhood feature and hidden representation of each node. SUGRL (Mo et al. 2022) designs a multiplet contrastive loss to guide positive pairs close while negative pairs far apart. In addition, the NT-Xent loss (Zhu et al. 2020) has been widely adopted by the state-of-the-art node-node GCL methods (You et al. 2020; Zhu et al. 2020; Wan et al. 2021a; Wan et al. 2021b; Zhu et al. 2021). Although the NT-Xent loss has been shown to be effective in CV, we argue that the definitions of positives and negatives might be inappropriate for GCL, since the graph data is not similar to image data. According to NT-Xent, the neighbors would be regarded as negatives and then pushed away from the anchor. However, this is undesirable in graph domain, as connected nodes are likely to share the same label and should not be far apart.

In NCLA, we propose a new neighbor contrastive loss for node-node GCL. Unlike NT-Xent, the proposed neighbor contrastive loss allows for multiple positives per anchor, which is similar to the supervised contrastive loss (Khosla et al. 2020). However, the supervised contrastive loss defines positives according to the observed class labels. While the proposed neighbor contrastive loss is designed for unsupervised GCL without access to class labels, instead, we take network topology as the supervised signals to define positives and negatives.

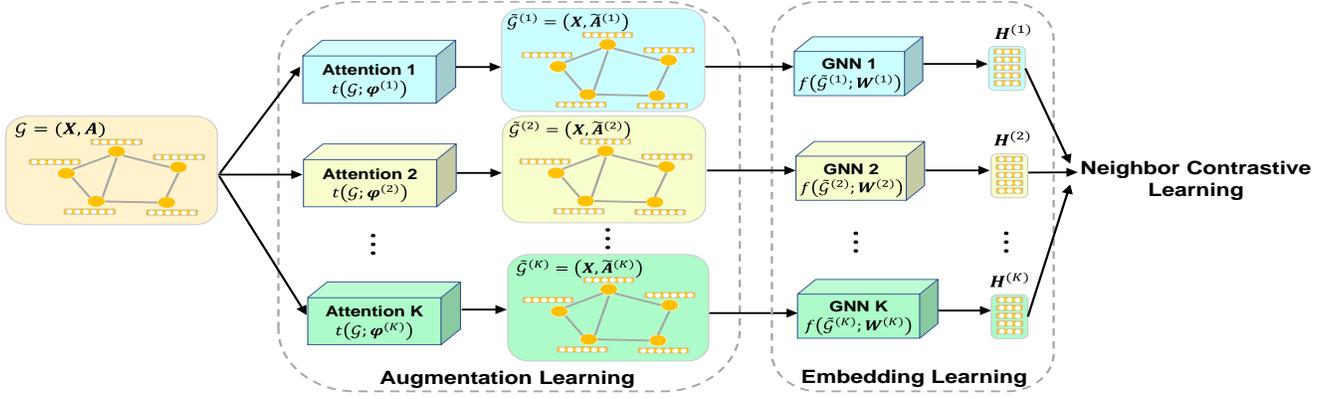

Figure 2. The model architecture of NCLA. It generates $K$ learnable augmented views with adaptive topology by multi-head GAT, where each $k$-th view has its own learnable parameters $\varphi^{(k)}, W^{(k)}$ which are not shared with other views. Then, it applies the neighbor contrastive loss to maximize the agreement between the embeddings of positive pairs and minimize that of negative pairs. Both augmentations and embeddings are learned end-to-end in NCLA.

## Methodology

In this section, we present the proposed NCLA in details, including how to generate learnable graph augmentation, how to define positives and negatives, and how to design the neighbor contrastive loss. Figure 2 shows the model architecture of NCLA.

### Preliminaries

Let $\mathcal{G} = (\mathcal{V}, \mathcal{E}, X, A)$ denote a graph, where $\mathcal{V} = \{v_1, v_2, \cdots, v_N\}$ and $\mathcal{E} \subseteq \mathcal{V} \times \mathcal{V}$ represent the node set and edge set respectively. $X \in \mathbb{R}^{N \times F}$ and $A \in \{0,1\}^{N \times N}$ denote the node feature matrix and the adjacency matrix, where $x_i \in \mathbb{R}^F$ is the feature vector of $v_i$ and $A_{ij} = 1$ iff $(v_i, v_j) \in \mathcal{E}$. $\mathcal{N}_i = \{v_j | j \neq i, A_{ij} = 1\}$ represents a set of first-order neighbors of $v_i$. Given $X$ and $A$ as the inputs, the proposed NCLA applies the augmentation function $t(\cdot; \varphi^{(k)})$ to generate the $k$-th learnable augmented view with adaptive topology $\tilde{\mathcal{G}}^{(k)} = t(\mathcal{G}; \varphi^{(k)}) = (X, \tilde{A}^{(k)}), k = 1, \cdots, K$, where $\tilde{A}^{(k)}$ is the adaptive adjacency matrix of the $k$-th augmented view and $K$ is the number of augmented views. Then, for each $k$-th augmented view $\tilde{\mathcal{G}}^{(k)}$, NCLA employs the GNN encoder $f(\cdot; W^{(k)})$ to learn the embeddings $H^{(k)} = f(\tilde{\mathcal{G}}^{(k)}; W^{(k)}) \in \mathbb{R}^{N \times F'}, F' \ll F$. The embeddings are learned by optimizing the graph contrastive loss, without access to the labels of downstream tasks.

### Learnable Graph Augmentation

In NCLA, we opt for multi-head GAT (Veličković et al. 2018) to generate $K$ learnable graph augmented views with adaptive topology $\{\tilde{A}^{(k)}\}_{k=1}^K$. Each $k$-th head, corresponding to the $k$-th augmentation function $t(\cdot; \varphi^{(k)})$, is constructed as a single-layer feedforward neural network. Specifically, the adaptive edge coefficient between two connected nodes, say $v_i$ and $v_j$, in the $k$-th augmented view can be learned as:

$$\tilde{A}_{ij}^{(k)} = \frac{e^{\text{LeakyReLU}(\varphi^{(k)}[W^{(k)}x_i \| W^{(k)}x_j])}}{\sum_{v_p \in \mathcal{N}_i \cup \{v_i\}} e^{\text{LeakyReLU}(\varphi^{(k)}[W^{(k)}x_i \| W^{(k)}x_p])}} \quad (1)$$

where $\tilde{A}_{ij}^{(k)}$ is set to 0 if $A_{ij} = 0$, $W^{(k)} \in \mathbb{R}^{F' \times F}$ is the learnable weight matrix to transform input features into the embedding space of the $k$-th augmented view, $\varphi^{(k)} \in \mathbb{R}^{1 \times 2F'}$ is the learnable weight vector of $k$-th head, $\|$ is the concatenation operation, and LeakyReLU$(\cdot)$ is a nonlinear activation function.

Then, for each $k$-th view, a GNN encoder $f(\cdot; W^{(k)})$ learns the embedding of each node by aggregating the neighbors' embeddings with adaptive edge coefficients and then applying the ELU nonlinearity, as:

$$h_i^{(k)} = \text{ELU}\left(\sum_{v_j \in \mathcal{N}_i \cup \{v_i\}} \tilde{A}_{ij}^{(k)} W^{(k)} x_j\right) \quad (2)$$

where $h_i^{(k)} \in \mathbb{R}^{F'}$ is the embedding of $v_i$ in the $k$-th view. Finally, the embeddings of each view are concatenated to generate the output embedding, as:

$$h_i = \|_{k=1}^K h_i^{(k)} \quad (3)$$

Compared to existing graph augmentations, the proposed attention-based learnable graph augmentation in NCLA has three advantages, as follows:

1) The hyper-parameters of handcraft graph augmentations have to be manually selected either randomly or by heuristics, which are sensitive to different graph datasets. In contrast, the end-to-end learnable graph augmentation in NCLA can be automatically adaptive to different graph datasets without human choices or prior domain knowledge.

2) The existing GCL methods usually adopt a shared-weight GNN encoder to learn the embeddings of different augmented views (You et al. 2020; Zhu et al. 2020; Zhu et al. 2021), which inevitably harms the diversity. In contrast, in NCLA, each $k$-th augmented view has its own learnable parameters of graph augmentation $\varphi^{(k)}$ and embedding learning $W^{(k)}$, which are distinct from other views. Thus, the diversity between different augmented views can be enhanced in NCLA.

3) The handcraft graph augmentations typically drop nodes or edges to generate different views, while dropping some sensitive elements can heavily damage the original graph topology which might be highly related to downstream tasks and consequently deteriorate the embedding quality (Zhu et al. 2021; Xia et al. 2022a). In contrast, NCLA learns adaptive coefficients of the original edges for different augmented views, thus yielding safer graph augmentation which enhances the diversity between different augmented views while avoiding improper modifications of original topology.

## Neighbor Contrastive Learning

GCL aims to maximize the mutual information (MI) between different augmented views, by contrasting positive pairs with negative counterparts. Two self-supervised contrastive losses, i.e., InfoNCE (Van den Oord, Li, and Vinyals 2018) and NT-Xent (Zhu et al. 2020), have been widely utilized in the latest node-node GCL methods (You et al. 2020; Zhu et al. 2020; Wan et al. 2021a; Wan et al. 2021b; Zhu et al. 2021). Both InfoNCE and NT-Xent only allow a single positive pair per anchor. Specifically, the embeddings of the same node in two views are defined as a positive pair, on the contrary, the embeddings of all different nodes are defined as negative pairs. In such definitions, the neighbors of an anchor would be treated as negatives, and then be pushed away from the anchor.

However, most GNNs are designed based on the homophily assumption which suggests that connected nodes tend to share similar labels and should be close to each other. Thus, we argue that the neighbors of the anchor should not be treated as negatives in GCL. To address this, we propose the concept of neighbor contrastive learning, which takes network topology as the supervised signals to define positives and negatives in node-node GCL. Specifically, not only the same node of the anchor in different views is treated as a positive, but also the neighbors of the anchor within a view and across different views would be treated as the extra positives. Besides, the non-neighbors of the anchor within a view and across different views would be treated as the negatives. The definitions of positive and negative pairs in NCLA are illustrated in Figure 1(c).

Let $h_i^{(1)}$ and $h_i^{(2)}$ denote the $L_2$-normalized embeddings of $v_i$ learned by view 1 and view 2 respectively. Selecting $h_i^{(1)}$ as the anchor, in NCLA, the positives come from three disjoint sources: 1) inter-view same node, i.e., the embedding of the same node in different view $h_i^{(2)}$; 2) intra-view neighbors, i.e., the embeddings of neighbors within a view $\left\{h_j^{(1)} | v_j \in \mathcal{N}_i\right\}$; and 3) inter-view neighbors, i.e., the embeddings of neighbors in different view $\left\{h_j^{(2)} | v_j \in \mathcal{N}_i\right\}$. That is, the number of positive pairs associated with the anchor $h_i^{(1)}$ should be $2|\mathcal{N}_i| + 1$, where $|\mathcal{N}_i|$ is the number of neighbors of $v_i$. The neighbor contrastive loss between view 1 and view 2 associated with the anchor $h_i^{(1)}$ is formulated as:

$$\ell(h_i^{(1)}) = -\log \frac{\left(e^{\theta(h_i^{(1)}, h_i^{(2)})/\tau} + \sum_{v_j \in \mathcal{N}_i}\left(e^{\theta(h_i^{(1)}, h_j^{(1)})/\tau} + e^{\theta(h_i^{(1)}, h_j^{(2)})/\tau}\right)\right)/(2|\mathcal{N}_i|+1)}{e^{\theta(h_i^{(1)}, h_i^{(2)})/\tau} + \sum_{j \neq i}\left(e^{\theta(h_i^{(1)}, h_j^{(1)})/\tau} + e^{\theta(h_i^{(1)}, h_j^{(2)})/\tau}\right)}$$

(4)

where $\tau$ is a temperature parameter, and $\theta(\cdot)$ denotes a similarity measure (here we use inner product). The last two terms in the denominator of Eq. (4) can be decomposed as:

$$\sum_{j \neq i} e^{\theta(h_i^{(1)}, h_j^{(1)})/\tau} = \underbrace{\sum_{v_j \in \mathcal{N}_i} e^{\theta(h_i^{(1)}, h_j^{(1)})/\tau}}_{\text{intra-view pos}} + \underbrace{\sum_{v_j \notin \mathcal{N}_i} e^{\theta(h_i^{(1)}, h_j^{(1)})/\tau}}_{\text{intra-view neg}}$$

$$\sum_{j \neq i} e^{\theta(h_i^{(1)}, h_j^{(2)})/\tau} = \underbrace{\sum_{v_j \in \mathcal{N}_i} e^{\theta(h_i^{(1)}, h_j^{(2)})/\tau}}_{\text{inter-view pos}} + \underbrace{\sum_{v_j \notin \mathcal{N}_i} e^{\theta(h_i^{(1)}, h_j^{(2)})/\tau}}_{\text{inter-view neg}}$$

where non-neighbors of $v_i$ in view 1 and view 2 are regarded as the intra-view and inter-view negatives respectively. Minimizing Eq. (4) would maximize the agreement between positive pairs and minimize that of negative pairs. Specifically, the embedding of each node would be forced to agree with itself in another view and its neighbors' embeddings within a view and across views. Oppositely, the embedding of each node can be distinguished from that of its non-neighbors within a view and across views. In addition, it is worth noting that the NT-Xent loss is a special case of the proposed neighbor contrastive loss, when only the inter-view same node is selected as a single positive, while the intra-view and inter-view neighbors are changed from positives to negatives.

Since two views are symmetric, given the embedding of $v_i$ in view 2, i.e., $h_i^{(2)}$ as the anchor, the neighbor contrastive loss $\ell(h_i^{(2)})$ can be similarly defined according to Eq. (4). The final neighbor contrastive loss between view 1 and view 2, averaged over all nodes is defined as:

$$\ell(H^{(1)}, H^{(2)}) = \frac{1}{2N} \sum_{i=1}^{N} \left[\ell(h_i^{(1)}) + \ell(h_i^{(2)})\right] \quad (5)$$

The proposed NCLA is flexible to generate arbitrary multiple learnable augmented views. When having more than two augmented views, we randomly choose one view as the

| Algorithm 1: NCLA |
| --- |
| **Input**: The adjacency matrix $A$, feature matrix $X$, number of augmented views $K$, number of training epochs $\mathcal{T}$. |
| 1     **for** epoch in 1 to $\mathcal{T}$ **do** |
| 2         **for** $k$ in 1 to $K$ **do** |
| 3             Generate the $k$-th learnable augmented views with Eq. (1); |
| 4             Generate embeddings of the $k$-th view with Eq. (2); |
| 5         **end for** |
| 6         Concatenate embeddings of $K$ views to generate final embeddings with Eq. (3); |
| 7         Compute neighbor contrastive loss $\mathcal{L}$ with Eq. (6); |
| 8         Update parameters by applying gradient descent to minimize $\mathcal{L}$; |
| 9     **end for** |
| 10    Generate embeddings with optimal parameters with Eqs. (1), (2) and (3). |
| **Output**: Embeddings $H$ |

pivot view, say $H^{(l)}$, and then sum up the neighbor contrastive loss between each other view and $H^{(l)}$ as the total neighbor contrastive loss, which is given by:

$$\mathcal{L} = \frac{1}{K}\sum_{k=1, k\neq l}^{K} \ell(H^{(k)}, H^{(l)}) \quad (6)$$

Note that both augmentations and embeddings are learned end-to-end in NCLA, which yields high level of flexibility and ease of use. The detailed description of NCLA is provided in Algorithm 1.

The time complexity of generating $K$ learnable augmented views by multi-head GAT is $O((NFF' + |\mathcal{E}|F')K)$, where $N$ and $|\mathcal{E}|$ are the number of nodes and edges in $\mathcal{G}$ respectively, $F$ is the number of input features and $F'$ is the embedding dimension. The time complexity of neighbor contrastive learning is $O(N^2F'(K-1))$. Thus, the time complexity of NCLA is $O((NFF' + |\mathcal{E}|F')K + N^2F'(K-1))$. Since $|\mathcal{E}| \ll N^2$, the overall time complexity of NCLA is $O((NFF' + N^2F')K)$. Note that in practice, $K$ is very small (e.g., 2 or 4) in our experiments, so the time complexity of NCLA is comparable to the representative node-node GCL methods, e.g. GRACE (Zhu et al. 2020) and GCA (Zhu et al. 2021).

## Experiments

### Datasets

To demonstrate the effectiveness of the proposed NCLA, extensive experiments have been conducted on five benchmark datasets for semi-supervised node classification, including three widely-used citation networks, i.e., Cora, Citeseer, Pubmed (Sen et al. 2008), a co-authorship network, i.e., Coauthor-CS (Shchur et al. 2018), and a product co-purchase network, i.e., Amazon-Photo (Shchur et al. 2018). The statistics of the datasets are summarized in Table 1.

### Baselines

NCLA is compared with 11 state-of-the-art methods for semi-supervised node classification, including 2 semi-supervised GNNs, i.e., **GCN** (Kipf and Welling 2017) and **GAT** (Veličković et al. 2018), 2 semi-supervised GCL methods, i.e., **CGPN** (Wan et al. 2021b) and **CG3** (Wan et al. 2021a), and 7 self-supervised GCL methods, i.e., **DGI** (Velickovic et al. 2019), **GMI** (Peng et al. 2020), **MVGRL** (Hassani and Khasahmadi 2020), **GRACE** (Zhu et al. 2020), **GCA** (Zhu et al. 2021), **SUGRL** (Mo et al. 2022) and **AFGRL** (Lee, Lee, and Park 2022).

### Experimental Settings

The proposed NCLA was implemented in PyTorch 1.10.1 (Paszke et al. 2019) and Deep Graph Library 0.6.1 (Wang et al. 2019). NCLA was trained by the Adam optimizer. The hyperparameters of NCLA on the five datasets are specified in Table 2. For the self-supervised GCL baselines and NCLA, the embeddings are learned in an unsupervised manner, and then used to train and test a $L_2$-regularized logistic regression (LR) classifier for semi-supervised node classification. On one hand, we conducted experiments in the setting with extremely scarce labels, where the number of training nodes per class $c$ was chosen in $\{1, 2, 3, 4\}$. In such a label-deficient setting, we followed (Li, Han, and Wu 2018; Li et al. 2019) to do not use a validation set with extra labels for model selection. On the other hand, we conducted experiments given relatively sufficient labels. For Cora, Citeseer and Pubmed, we followed (Yang, Cohen, and Salakhudinov 2016) to randomly select 20 nodes per class for training, 500 nodes for validation and the remaining nodes for test. For Coauthor CS and Amazon Photo, we followed (Liu, Gao, and Ji 2020) to randomly select 20 nodes per class for training, 30 nodes per class for validation, and the remaining nodes for test. Note that for the self-supervised GCL baselines and NCLA which learn embeddings from unlabeled data, the validation set was just used to tune the hyperparameters of the LR classifier, rather than the GCL models. For each dataset, we conducted 20 random splits of training/validation/test, and reported the averaged performance of all algorithms on the same random splits.

| Datasets | # Nodes | # Edges | # Features | # Labels |
| --- | --- | --- | --- | --- |
| Cora | 2708 | 10556 | 1433 | 7 |
| CiteSeer | 3327 | 9228 | 3703 | 6 |
| PubMed | 19717 | 88651 | 500 | 3 |
| Coauthor-CS | 18333 | 163788 | 6805 | 15 |
| Amazon-Photo | 7650 | 238162 | 745 | 8 |

Table 1: Statistics of the datasets.

| Datasets | # Augmented Views $K$ | # Hidden Layers | # Embedding Dimension $F'$ | Temperature $\tau$ | Learning Rate | Weight Decay | # Epochs $\mathcal{T}$ |
|---|---|---|---|---|---|---|---|
| Cora | 4 | 1 | 32 | 1 | 1e-2 | 1e-4 | 2000 |
| CiteSeer | 4 | 1 | 32 | 5 | 1e-2 | 1e-4 | 2000 |
| PubMed | 2 | 1 | 32 | 5 | 1e-3 | 5e-5 | 2000 |
| Coauthor-CS | 4 | 1 | 32 | 1 | 5e-2 | 1e-4 | 2000 |
| Amazon-Photo | 2 | 1 | 32 | 1 | 1e-3 | 1e-4 | 2000 |

Table 2: Hyperparameter Settings of NCLA.

| Datasets | c | GCN | GAT | CGPN | CG3 | DGI | GMI | MVGRL | GRACE | GCA | SUGRL | AFGRL | NCLA |
|---|---|---|---|---|---|---|---|---|---|---|---|---|---|
| Cora | 1 | 42.6±11.6 | 42.1±9.5 | 58.6±10.6 | 55.4±14.3 | 55.4±11.4 | 55.9±9.6 | 59.1±10.9 | 51.0±9.8 | 58.4±10.9 | 55.2±8.9 | 47.7±7.8 | **63.1±11.2** |
| | 2 | 55.0±7.5 | 53.2±9.0 | 67.4±7.1 | 66.4±7.7 | 64.9±9.0 | 65.2±7.6 | 67.8±8.6 | 59.7±7.9 | 66.0±7.8 | 65.3±6.2 | 57.8±6.8 | **71.8±6.9** |
| | 3 | 63.1±6.8 | 63.2±5.3 | 70.7±4.0 | 71.5±4.2 | 71.1±5.6 | 70.7±5.2 | 74.5±4.1 | 64.0±6.6 | 71.5±4.6 | 70.5±3.5 | 64.6±4.7 | **75.7±5.0** |
| | 4 | 66.4±6.4 | 66.3±5.9 | 70.7±2.9 | 72.7±2.4 | 72.9±4.5 | 73.3±4.3 | 76.1±3.2 | 66.1±5.4 | 72.9±4.3 | 73.5±2.9 | 67.5±4.2 | **77.3±3.8** |
| | 20 | 79.6±1.8 | 81.2±1.6 | 74.0±1.7 | 80.6±1.6 | 82.1±1.3 | 79.4±1.2 | **82.4±1.5** | 79.6±1.4 | 79.0±1.4 | 81.3±1.2 | 78.6±1.3 | 82.2±1.6 |
| Citeseer | 1 | 33.8±5.9 | 31.0±7.2 | 48.6±11.3 | 48.4±12.8 | 47.2±9.2 | 40.8±6.8 | 32.8±8.4 | 40.3±7.2 | 38.7±9.0 | 46.7±8.4 | 42.1±7.2 | **52.2±13.5** |
| | 2 | 44.8±5.5 | 41.1±7.2 | 58.0±5.1 | 60.2±6.8 | 58.6±4.3 | 50.2±4.1 | 47.8±7.5 | 48.5±6.0 | 49.6±5.3 | 57.7±4.6 | 53.3±5.4 | **62.2±6.4** |
| | 3 | 49.2±5.1 | 48.6±6.7 | 59.4±5.4 | 62.1±7.3 | 63.3±4.3 | 55.1±2.7 | 55.2±6.7 | 52.7±4.6 | 54.2±4.7 | 61.8±5.1 | 58.0±4.4 | **65.5±3.5** |
| | 4 | 51.7±4.5 | 52.8±6.6 | 60.6±3.4 | 65.1±2.5 | 65.8±2.1 | 57.9±3.0 | 59.3±5.5 | 56.0±3.9 | 57.3±3.3 | 65.0±2.6 | 61.5±2.5 | **67.6±2.1** |
| | 20 | 66.0±1.2 | 68.9±1.8 | 63.7±1.6 | 70.9±1.5 | 71.6±1.2 | 66.9±2.2 | 71.1±1.4 | 67.0±1.7 | 65.6±2.4 | 71.0±1.8 | 70.8±2.1 | **71.7±0.9** |
| PubMed | 1 | 48.6±7.1 | 47.9±8.5 | 53.5±13.4 | 54.7±8.6 | 50.0±9.5 | 53.5±11.9 | 55.3±9.3 | 46.5±7.0 | 57.7±10.5 | 56.7±8.8 | 49.7±8.3 | **60.2±12.4** |
| | 2 | 55.8±7.1 | 54.5±7.7 | 59.7±10.3 | 58.9±7.2 | 58.5±8.7 | 60.7±9.9 | 62.7±7.0 | 53.8±6.9 | 66.3±7.6 | 62.9±6.3 | 56.4±6.4 | **66.9±9.7** |
| | 3 | 62.1±7.3 | 61.5±6.8 | 61.8±10.4 | 65.1±6.5 | 62.4±7.2 | 65.5±8.9 | 68.5±5.8 | 55.6±7.9 | 71.9±5.4 | 67.9±5.7 | 60.6±5.5 | **72.3±6.2** |
| | 4 | 65.1±5.9 | 64.2±6.1 | 62.7±10.3 | 66.0±5.7 | 64.1±6.2 | 67.2±8.1 | 70.6±6.0 | 57.7±6.8 | 73.6±5.4 | 69.9±5.1 | 62.4±5.1 | **73.8±4.9** |
| | 20 | 79.0±2.5 | 78.5±1.8 | 73.3±2.5 | 78.9±2.6 | 78.3±2.4 | 76.8±2.3 | 79.5±2.2 | 74.6±3.5 | 81.5±2.5 | 80.5±1.6 | 76.4±2.5 | **82.0±1.4** |
| Coauthor CS | 1 | 64.8±8.8 | 64.2±9.0 | 68.4±8.9 | 79.8±8.0 | 71.4±6.3 | 68.3±7.2 | 75.4±7.2 | 60.0±7.7 | 59.9±7.6 | 76.9±6.2 | 75.2±7.6 | **83.0±6.2** |
| | 2 | 79.2±4.2 | 80.2±4.1 | 77.7±5.3 | 85.3±4.0 | 79.6±5.3 | 78.1±4.5 | 84.7±2.7 | 71.3±4.5 | 72.5±4.6 | 85.4±3.1 | 85.3±2.7 | **87.4±4.1** |
| | 3 | 83.3±4.0 | 85.0±2.7 | 80.4±4.4 | 87.5±3.9 | 82.3±3.6 | 80.9±4.4 | 87.5±2.2 | 74.8±3.8 | 77.9±4.1 | 87.4±2.9 | 87.7±2.3 | **88.3±3.1** |
| | 4 | 84.2±3.1 | 86.6±2.1 | 80.9±3.6 | 87.1±4.6 | 84.8±2.8 | 82.8±2.8 | 88.5±1.8 | 77.6±2.8 | 80.3±3.1 | 88.2±2.1 | 88.4±1.9 | **88.8±2.4** |
| | 20 | 90.0±0.6 | 90.9±0.7 | 83.5±1.4 | 90.6±1.0 | **92.0±0.5** | 88.5±0.8 | 91.5±0.6 | 90.0±0.7 | 90.9±1.1 | 91.2±0.9 | 91.4±0.6 | 91.5±0.7 |
| Amazon Photo | 1 | 60.7±9.3 | 59.0±11.5 | 70.4±7.2 | 69.3±5.8 | 53.8±10.7 | 58.2±8.1 | 59.7±9.0 | 67.0±9.0 | 55.3±6.7 | 71.6±6.2 | 54.4±9.9 | **75.6±6.0** |
| | 2 | 75.2±7.2 | 71.7±6.4 | 75.7±4.3 | 77.2±3.6 | 62.7±8.5 | 68.8±6.2 | 73.4±6.8 | 76.6±5.2 | 68.0±5.6 | 80.7±3.6 | 71.3±7.2 | **81.6±3.7** |
| | 3 | 76.9±5.1 | 75.6±6.3 | 77.0±4.0 | 79.4±3.9 | 66.7±6.7 | 71.9±5.4 | 76.8±6.1 | 78.6±4.8 | 74.4±5.9 | 82.2±2.7 | 75.9±5.7 | **83.3±3.8** |
| | 4 | 81.0±4.6 | 79.3±5.9 | 80.1±2.6 | 81.9±2.9 | 70.8±6.0 | 76.2±1.8 | 82.0±2.3 | 81.8±1.4 | 78.8±3.9 | 84.3±1.6 | 81.5±2.5 | **85.3±2.0** |
| | 20 | 86.3±1.6 | 86.5±2.1 | 84.1±1.5 | 89.4±1.9 | 83.5±1.2 | 86.7±1.5 | 89.7±1.2 | 87.9±1.4 | 87.0±1.9 | **90.5±1.9** | 89.2±1.1 | 90.2±1.3 |

Table 3: Classification accuracy with different label rates on five datasets. The best and second-best results are highlighted in boldface and underlined respectively.

## Node Classification Results

Table 3 shows the node classification accuracy on five benchmark graph datasets. We have the following observations:

Firstly, NCLA substantially outperforms GAT by a large margin, e.g., NCLA improves on GAT by 21%, when 1 labeled node per class is used for training on Cora. Recall that NCLA utilizes multi-head GAT as the backbone to generate the attention-based learnable graph augmentation. The significant improvement of NCLA over multi-head GAT reflects that more robust representations can be learned with the help of GCL, especially when the labels are extremely limited.

Moreover, NCLA consistently outperforms the state-of-the-art GCL baselines on all datasets when the labels are extremely limited (i.e., only 1, 2, 3 and 4 labeled nodes per class). When more labels are given, i.e., 20 labeled nodes per class, NCLA achieves the best or second-best results, which are comparable to previous state-of-the-art. The outperformance of NCLA is mainly attributed to two folds. On one hand, unlike the GCL baselines which have to manually tune handcraft augmentations per dataset, NCLA generates the attention-based learnable augmentation along with embeddings end-to-end. The inappropriate handcraft augmentations might heavily damage the network topology and lead to ineffective embeddings. In contrast, NCLA brings safer learnable graph augmentation which avoids inappropriate modification of the original topology. On the other hand, the

| Variants | Cora | CiteSeer | PubMed | CS | Photo |
|---|---|---|---|---|---|
| NCL* | **63.1** | **52.2** | **60.2** | **83.0** | **75.6** |
| InfoNCE | 60.9 | 48.5 | 59.7 | 80.4 | 75.3 |
| NT-Xent | 60.1 | 51.6 | 60.0 | 80.5 | 75.2 |
| NCL* w/o Pos 2 | 62.4 | 51.0 | 59.6 | 82.7 | 72.5 |
| NCL* w/o Pos 3 | 62.3 | 49.5 | 59.4 | 81.4 | 71.2 |

*NCL is short for neighbor contrastive loss.

Table 4. Variants of neighbor contrastive loss in NCLA.

GCL baselines generally adopt the contrastive losses in CV, without considering the network topology. As a result, the neighbors would be treated as negatives and then pushed away from the anchor. In contrast, the proposed neighbor contrastive loss in NCLA employs the neighbors of the anchor within a view and from different views as two extra sources of positives, thus making each node not only agree with itself in different augmented views, but also agree with its intra-view and inter-view neighbors. Under the label-deficient setting, taking full advantage of network topology can propagate the scarce labeled information through the neighborhood to effectively boost the node classification performance (Li et al. 2019). Thus, the proposed neighbor contrastive loss which takes network topology as the supervised signals to define positives and negatives in GCL can demonstrate more effectiveness, especially when the labels are extremely limited.

### Ablation Study

Next, we investigate the variants of the proposed neighbor contrastive loss. As illustrated in Figure 1, NT-Xent is a special case of the proposed neighbor contrastive loss by changing the intra-view neighbors and inter-view neighbors from positives to negatives. InfoNCE is a special case of NT-Xent by removing intra-view different nodes from negatives. Additionally, we study two more variants of the proposed neighbor contrastive loss, by changing either intra-view neighbors (i.e. Pos 2) or inter-view neighbors (i.e. Pos 3) from positives to negatives. As shown in Table 4, the proposed neighbor contrastive loss consistently yields the highest accuracy among all the loss variants on the five datasets. Compared to InfoNCE and NT-Xent, the proposed neighbor contrastive loss achieves significant gains on Cora, Citeseer and Coauthor CS, and comparable performance on PubMed and Amazon Photo. In addition, without either intra-view or inter-view neighbors as positives would lead to worse results. This reflects that selecting the neighbors within a view and from different views as positives are both useful for neighbor contrastive learning in NCLA.

### Hyperparameter Analysis

Figure 3 shows the sensitivity analysis on the hyperparameters $K$, $F'$ and $\tau$ of NCLA on two various graph datasets, i.e.,

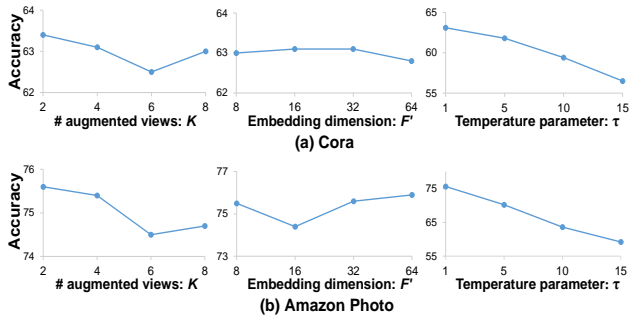

Figure 3. Sensitivity analysis of the hyperparameters $K$, $F'$ and $\tau$ on NCLA.

Cora (citation network) and Amazon Photo (e-commerce network). We observe that the number of augmented views $K$=2 yields the best performance, while $K$=6 leads to the worst results on both datasets. Actually, more augmented views lead to higher complexity, thus, it is suggested to generate 2 or 4 views in NCLA. The performance of NCLA is stable on various embedding dimensions $F'$ in $\{8, 16, 32\}$ on Cora, while $F'$=16 leads to the worst result on Amazon Photo. The temperature parameter $\tau$ of 1 achieves the best performance while larger $\tau$ leads to lower accuracy on both datasets.

## Conclusion

Despite the prosperous development of GCL, how to generate learnable graph augmentation and devise the contrastive loss more suitable for GCL remains rarely explored. To fill in this gap, we propose an end-to-end automatic GCL method, named NCLA. On one hand, NCLA employs multi-head GAT to generate the attention-based learnable graph augmentation, which can be automatically adaptive to various graph datasets. Each augmented view consists of exactly the same nodes and edges as the original graph but with different adaptive edge weights. In addition, each view has its own learnable parameters for both graph augmentation and embedding learning. As a result, the diversity between different augmented views can be enhanced, while avoiding improper modification of the original topology. On the other hand, instead of directly utilizing the contrastive losses in CV which overlook network topology, NCLA proposes a new neighbor contrastive loss to allow for multiple positives per anchor, by taking network topology as the supervised signals. Specifically, each anchor forms positive pairs with not only the same node in different view, but also its neighbors within a view and across different views. Both augmentations and embeddings are learned end-to-end in NCLA. The extensive node classification experiments demonstrate that NCLA can consistently gain the superior results compared to the state-of-the-art GCL methods or even some supervised GNNs, when the labels are extremely limited.

## Acknowledgments

This work was supported in part by National Natural Science Foundation of China (No. 62102124), Hainan Provincial Natural Science Foundation of China (No. 322RC570), and the Research Start-up Fund of Hainan University (No. KYQD(ZR)-22016).

## References

Aberdam, A.; Litman, R.; Tsiper, S.; Anschel, O.; Slossberg, R.; Mazor, S.; Manmatha, R.; and Perona, P. 2021. Sequence-to-sequence Contrastive Learning for Text Recognition. In *Proceedings of the IEEE/CVF Conference on Computer Vision and Pattern Recognition*, 15302-15312.

Dai, Q.; Wu, X.-M.; Xiao, J.; Shen, X.; and Wang, D. 2022. Graph Transfer Learning via Adversarial Domain Adaptation with Graph Convolution. *IEEE Transactions on Knowledge and Data Engineering*.

Ding, K.; Xu, Z.; Tong, H.; and Liu, H. 2022. Data Augmentation for Deep Graph Learning: A Survey. *ACM SIGKDD Explorations Newsletter* 24(2): 61-77.

Hamilton, W.; Ying, Z.; and Leskovec, J. 2017. Inductive Representation Learning on Large Graphs. In *Proceedings of Advances in Neural Information Processing Systems*, 1024-1034.

Hassani, K., and Khasahmadi, A. H. 2020. Contrastive Multi-view Representation Learning on Graphs. In *Proceedings of International Conference on Machine Learning*, 4116-4126.

Khosla, P.; Teterwak, P.; Wang, C.; Sarna, A.; Tian, Y.; Isola, P.; Maschinot, A.; Liu, C.; and Krishnan, D. 2020. Supervised Contrastive Learning. In *Proceedings of Advances in Neural Information Processing Systems*, 18661-18673.

Kipf, T. N., and Welling, M. 2017. Semi-supervised Classification with Graph Convolutional Networks. In *Proceedings of International Conference on Learning Representations*.

Lee, N.; Lee, J.; and Park, C. 2022. Augmentation-free Self-supervised Learning on Graphs. In *Proceedings of AAAI Conference on Artificial Intelligence*.

Li, Q.; Han, Z.; and Wu, X.-M. 2018. Deeper Insights into Graph Convolutional Networks for Semi-supervised Learning. In *Proceedings of AAAI Conference on Artificial Intelligence*, 3538-3545.

Li, Q.; Wu, X.-M.; Liu, H.; Zhang, X.; and Guan, Z. 2019. Label Efficient Semi-supervised Learning via Graph Filtering. In *Proceedings of the IEEE/CVF Conference on Computer Vision and Pattern Recognition*, 9582-9591.

Linsker, R. 1988. Self-organization in a Perceptual Network. *Computer* 21(3): 105-117.

Liu, M.; Gao, H.; and Ji, S. 2020. Towards Deeper Graph Neural Networks. In *Proceedings of the 26th ACM SIGKDD International Conference on Knowledge Discovery & Data Mining*, 338-348.

Liu, Y.; Jin, M.; Pan, S.; Zhou, C.; Zheng, Y.; Xia, F.; and Yu, P. 2022. Graph Self-supervised Learning: A Survey. *IEEE Transactions on Knowledge and Data Engineering*.

McPherson, M.; Smith-Lovin, L.; and Cook, J. M. 2001. Birds of a Feather: Homophily in Social Networks. *Annual Review of Sociology*: 415-444.

Mo, Y.; Peng, L.; Xu, J.; Shi, X.; and Zhu, X. 2022. Simple Unsupervised Graph Representation Learning. In *Proceedings of AAAI Conference on Artificial Intelligence*.

Nowozin, S.; Cseke, B.; and Tomioka, R. 2016. F-GAN: Training Generative Neural Samplers using Variational Divergence Minimization. In *Proceedings of Advances in Neural Information Processing Systems*, 271-279.

Paszke, A.; Gross, S.; Massa, F.; Lerer, A.; Bradbury, J.; Chanan, G.; Killeen, T.; Lin, Z.; Gimelshein, N.; and Antiga, L. 2019. Pytorch: An Imperative Style, High-performance Deep Learning Library. In *Proceedings of Advances in Neural Information Processing Systems*, 8026-8037.

Peng, Z.; Huang, W.; Luo, M.; Zheng, Q.; Rong, Y.; Xu, T.; and Huang, J. 2020. Graph Representation Learning via Graphical Mutual Information Maximization. In *Proceedings of The Web Conference 2020*, 259-270.

Qiu, J.; Chen, Q.; Dong, Y.; Zhang, J.; Yang, H.; Ding, M.; Wang, K.; and Tang, J. 2020. GCC: Graph Contrastive Coding for Graph Neural Network Pre-training. In *Proceedings of the 26th ACM SIGKDD International Conference on Knowledge Discovery & Data Mining*, 1150-1160.

Schroff, F.; Kalenichenko, D.; and Philbin, J. 2015. Facenet: A Unified Embedding for Face Recognition and Clustering. In *Proceedings of the IEEE Conference on Computer Vision and Pattern Recognition*, 815-823.

Sen, P.; Namata, G.; Bilgic, M.; Getoor, L.; Galligher, B.; and Eliassi-Rad, T. 2008. Collective Classification in Network Data. *AI Magazine* 29(3): 93-93.

Shchur, O.; Mumme, M.; Bojchevski, A.; and Günnemann, S. 2018. Pitfalls of Graph Neural Network Evaluation. *arXiv preprint arXiv:1811.05868*.

Shen, X., and Chung, F.-L. 2020. Deep Network Embedding for Graph Representation Learning in Signed Networks. *IEEE Transactions on Cybernetics* 50(4): 1556-1568. DOI: 10.1109/TCYB.2018.2871503.

Shen, X.; Dai, Q.; Chung, F.-l.; Lu, W.; and Choi, K.-S. 2020a. Adversarial Deep Network Embedding for Cross-network Node Classification. In *Proceedings of the AAAI Conference on Artificial Intelligence*, 2991-2999.

Shen, X.; Dai, Q.; Mao, S.; Chung, F.-l.; and Choi, K.-S. 2020b. Network Together: Node Classification via Cross-network Deep Network Embedding. *IEEE Transactions on Neural Networks and Learning Systems* 32(5): 1935-1948.

Shen, X.; Mao, S.; and Chung, F.-l. 2020. Cross-network Learning with Fuzzy Labels for Seed Selection and Graph Sparsification in Influence Maximization. *IEEE*

*Transactions on Fuzzy Systems* 28(9): 2195-2208. DOI: 10.1109/TFUZZ.2019.2931272.

Tian, Y.; Sun, C.; Poole, B.; Krishnan, D.; Schmid, C.; and Isola, P. 2020. What Makes for Good Views for Contrastive Learning? In *Proceedings of Advances in Neural Information Processing Systems*, 6827-6839.

Van den Oord, A.; Li, Y.; and Vinyals, O. 2018. Representation Learning with Contrastive Predictive Coding. *arXiv e-prints*: arXiv: 1807.03748.

Veličković, P.; Cucurull, G.; Casanova, A.; Romero, A.; Liò, P.; and Bengio, Y. 2018. Graph Attention Networks. In *Proceedings of International Conference on Learning Representations*.

Velickovic, P.; Fedus, W.; Hamilton, W. L.; Liò, P.; Bengio, Y.; and Hjelm, R. D. 2019. Deep Graph Infomax. In *Proceedings of International Conference on Learning Representations*.

Wan, S.; Pan, S.; Yang, J.; and Gong, C. 2021a. Contrastive and Generative Graph Convolutional Networks for Graph-based Semi-supervised Learning. In *Proceedings of AAAI Conference on Artificial Intelligence*.

Wan, S.; Zhan, Y.; Liu, L.; Yu, B.; Pan, S.; and Gong, C. 2021b. Contrastive Graph Poisson Networks: Semi-supervised Learning with Extremely Limited Labels. In *Proceedings of Advances in Neural Information Processing Systems*.

Wang, H.; Zhang, J.; Zhu, Q.; and Huang, W. 2022. Augmentation-free Graph Contrastive Learning. *arXiv preprint arXiv:2204.04874*.

Wang, M.; Zheng, D.; Ye, Z.; Gan, Q.; Li, M.; Song, X.; Zhou, J.; Ma, C.; Yu, L.; and Gai, Y. 2019. Deep Graph Library: A Graph-centric, Highly-performant Package for Graph Neural Networks. *arXiv preprint arXiv:1909.01315*.

Wu, M.; Pan, S.; Zhou, C.; Chang, X.; and Zhu, X. 2020. Unsupervised Domain Adaptive Graph Convolutional Networks. In *Proceedings of The Web Conference 2020*, 1457-1467.

Xia, J.; Wu, L.; Chen, J.; Hu, B.; and Li, S. Z. 2022a. SimGRACE: A Simple Framework for Graph Contrastive Learning without Data Augmentation. In *Proceedings of the ACM Web Conference 2022*, 1070-1079.

Xia, J.; Wu, L.; Wang, G.; and Li, S. Z. 2022b. ProGCL: Rethinking Hard Negative Mining in Graph Contrastive Learning. In *Proceedings of International Conference on Machine Learning*.

Yang, H.; Chen, H.; Pan, S.; Li, L.; Yu, P. S.; and Xu, G. 2022. Dual Space Graph Contrastive Learning. In *Proceedings of the ACM Web Conference 2022*, 1238-1247.

Yang, Z.; Cohen, W.; and Salakhudinov, R. 2016. Revisiting Semi-supervised Learning with Graph Embeddings. In *Proceedings of International Conference on Machine Learning*, 40-48.

You, Y.; Chen, T.; Sui, Y.; Chen, T.; Wang, Z.; and Shen, Y. 2020. Graph Contrastive Learning with Augmentations. In *Proceedings of Advances in Neural Information Processing Systems*, 5812-5823.

Zheng, Y.; Pan, S.; Lee, V. C.; Zheng, Y.; and Yu, P. S. 2022. Rethinking and Scaling Up Graph Contrastive Learning: An Extremely Efficient Approach with Group Discrimination. In *Proceedings of Advances in Neural Information Processing Systems*.

Zhu, Y.; Xu, Y.; Yu, F.; Liu, Q.; Wu, S.; and Wang, L. 2020. Deep Graph Contrastive Representation Learning. In *Proceedings of ICML Workshop on Graph Representation Learning and Beyond*.

Zhu, Y.; Xu, Y.; Yu, F.; Liu, Q.; Wu, S.; and Wang, L. 2021. Graph Contrastive Learning with Adaptive Augmentation. In *Proceedings of the Web Conference 2021*, 2069-2080.